\documentclass[12pt]{iopart}
\usepackage{txfonts,bm,pifont}  
\usepackage{cite}  
\usepackage{graphicx,xcolor}
\usepackage{amssymb}
\usepackage{iopams}

\newcommand{\taus}{\tau^*{}}
\def\journal#1&#2,{\begingroup \let\journal=\dummyjournal
               \it #1\unskip~\bf\ignorespaces #2\rm,\endgroup}
\def\dummyjournal{\errmessage{Reference foul up: nested \journal macros}}
\makeatletter
\@addtoreset{equation}{section}
\renewcommand{\theequation}{\@arabic\c@section.\@arabic\c@equation}
\long\def\@makecaption#1#2{
 \vskip 10pt 
 \setbox\@tempboxa\hbox{#1. #2}
 \ifdim \wd\@tempboxa >\hsize #1. #2\par \else \hbox
to\hsize{\hfil\box\@tempboxa\hfil} 
 \fi}
\makeatother
\begin{document}
\title[Discrete Integrable Systems and Hodograph Transformations]{Discrete Integrable Systems and
Hodograph Transformations Arising from Motions of Discrete Plane Curves }

\author{Bao-Feng Feng$^1$, Jun-ichi Inoguchi$^2$, 
Kenji Kajiwara$^{3}$, Ken-ichi Maruno
$^1$ and 
Yasuhiro Ohta$^{4}$ 
}
\address{$^1$~Department of Mathematics,
The University of Texas-Pan American,
Edinburg, TX 78539-2999
}
\address{
$^2$~Department of Mathematical Sciences, Yamagata University,
 1-4-12 Kojirakawa-machi, Yamagata 990-8560, Japan}
\address{$^3$~Institute of Mathematics for Industry, Kyushu University, 744 Motooka, Fukuoka 819-8581, Japan
}
\address{$^4$~Department of Mathematics, Kobe University, Rokko, Kobe
  657-8501, Japan
}
\date{\today}
\def\submitto#1{\vspace{28pt plus 10pt minus 18pt}
     \noindent{\small\rm To be submitted to : {\it #1}\par}}

\begin{abstract}
We consider integrable discretizations of some soliton equations associated with the motions of
plane curves: the Wadati-Konno-Ichikawa elastic beam equation, the complex Dym equation, and the short pulse
equation. They are related to the modified KdV or the sine-Gordon equations by the hodograph
 transformations.  Based on the observation that the hodograph transformations are regarded as the
Euler-Lagrange transformations of the curve motions, we construct the discrete analogues of the hodograph
transformations, which yield integrable discretizations of 
those soliton equations.
 \par \kern\bigskipamount\noindent \today
\end{abstract}

\kern-\bigskipamount

\submitto{\JPA}

\section{Introduction}

The study of discrete integrable systems has received considerable attention in the past decade
(see, for example, \cite{Grammaticos:book}). 
Ablowitz and Ladik proposed a method of integrable
discretizations of soliton equations, including the nonlinear Schr\"odinger equation and the
modified KdV (mKdV) equation, based on the Lax form~\cite{Ablowitz-Ladik:1975,Ablowitz-Ladik:1977,Ablowitz:book}, and Hirota discretized
various soliton equations such as the KdV, the mKdV, and the sine-Gordon equations by using the
bilinear
formalism\cite{Hirota:difference1,Hirota:difference2,Hirota:dsG,Hirota:difference4,Hirota:difference5}.
Following 
the pioneering work of Ablowitz-Ladik and Hirota, Date, Jimbo and Miwa developed a unified
algebraic approach from the point of view of the KP
theory\cite{DJM:discrete1,DJM:discrete2,DJM:discrete3,DJM:discrete4,DJM:discrete5,Jimbo-Miwa,Miwa}.
For other approaches to the discrete integrable systems, see, for example,
\cite{Suris:book,Nijhoff}.

It is known that there is a class of soliton equations which admits loop, cusp, and peak soliton
solutions.  Among them, some soliton equations, e.g. the Wadati-Konno-Ichikawa (WKI) elastic beam
equation, the Camassa-Holm equation, the Degasperis-Procesi equation, the short pulse equation and
the Hunter-Saxton equation, appear as model equations of various physical
phenomena~\cite{WKI1,WKI2,WKI3,Konno,CH,DP,SP,HS}. 
It should be noted that those equations are
transformed to certain soliton equations which admit smooth soliton solutions through hodograph
transformations~\cite{Ishimori,Ishimori2,
Shiff-CH,Matsuno-CH,Matsuno-DP1,Matsuno-DP2,
Matsuno-SP,Matsuno-HS,Rogers:book}. 
For example, the WKI elastic beam equation is transformed to the potential mKdV
equation~\cite{Ishimori}.

Integrable discretization of those soliton equations 
has been regarded as a difficult problem until
recently. Especially, a systematic treatment of hodograph transformations in discretizations has
been unknown. 
Recently, some of the authors succeeded in integrable discretization of some
equations in the above class of soliton equations by using the bilinear
method, and it was confirmed that those integrable discrete equations 
work effectively on numerical 
computations of the above class of soliton equations 
as self-adaptive mesh schemes~
\cite{discreteCH,discreteSP,discreteHS,selfadaptive}.
However, the method employed there is rather technical, so it is not easy to extract a fundamental
structure of discretizations to apply this method to a broader 
class of nonlinear wave equations.

On the other hand, the discrete integrable systems have been applied to discretizations of curves
and surfaces, and this area has been recently studied actively under the name of the discrete
differential geometry~\cite{Bobenko:book1,Bobenko:book2}.  In particular, there have been intensive
studies in topics related to curve geometry after the pioneering work of Lamb and
Goldstein-Petrich~\cite{Lamb,Goldstein-Petrich}, and then several 
frameworks for the motion of
discrete curves have been proposed in various 
settings~\cite{DS1,DS2,DS3,HK,Inoguchi,Matsuura}.

It is well known that the potential mKdV equation describes the motion of plane curves~\cite{Goldstein-Petrich}.
Recently, the authors considered continuous and discrete motion of 
discrete plane curves in the
Euclidean plane and presented the explicit formula in terms of the
$\tau$-function~\cite{Inoguchi,Matsuura}. 
The hodograph transformation of the WKI elastic beam
equation can be viewed as the Euler-Lagrange transformation of the motion of plane curves. From this
fact, it may be possible to establish a discrete analogue of the hodograph transformation for the
motion of discrete curves, and to discretize soliton equations having singularities naturally from a
geometric point of view.

In the present paper, we construct discrete analogues of hodograph transformations by the
Euler-Lagrange transformations of the motion of discrete plane 
curves in the Euclidean plane.  Based
on them, we construct semi-discrete and fully discrete analogues of the WKI elastic beam equation,
the complex Dym equation, and the short pulse equation.  In Section 2.1, we discuss the motion of
plane curves described by the potential mKdV equation and the hodograph transformations for the WKI
elastic beam equation and the complex Dym equation.  We also discuss the relationship between the
sine-Gordon equation and the short pulse equation.

In Section 2.2, we introduce the discrete hodograph transformation for the continuous motion of
discrete plane curves which are described by the semi-discrete potential mKdV equation.  Then we construct the
semi-discrete WKI elastic beam equation and the semi-discrete complex Dym equation.  Using the same
technique to the semi-discrete sine-Gordon equation, we construct the semi-discrete short pulse
equation.

In Section 2.3, we consider the discrete motion of discrete plane 
curves which are described by the
discrete potential mKdV equation.  Then we construct the discrete WKI elastic beam equation and the
discrete complex Dym equation by using the discrete hodograph transformations.  In a similar way, we
construct the discrete short pulse equation from the discrete sine-Gordon equation.

In Section 3, we present a $\tau$-function which gives soliton and breather solutions for these
equations, and Hirota-type bilinear equations for the $\tau$-function. Section 4 is devoted to
the conclusion.

\section{Motion of plane curves and hodograph transformations}
\subsection{Motion of smooth curves}
Let $\gamma(s)$ be an arc-length parametrized curve in Euclidean plane $\mathbb{R}^2$. Then the
tangent vector $\frac{\partial\gamma}{\partial s}$ satisfies
\begin{equation}
 \left|\frac{\partial\gamma}{\partial s}\right|=1. \label{gamma':isoperimetricity}
\end{equation}
Thus $\frac{\partial\gamma}{\partial s}$ admits the parametrization 
\begin{equation}
\frac{\partial\gamma}{\partial s}=\left[\begin{array}{l}\cos\theta\\\sin\theta\end{array}\right].
\label{angle_function}
\end{equation}
The function $\theta=\theta(s)$ is called the angle function of $\gamma$ which denotes the angle of
$\frac{\partial\gamma}{\partial s}$ measured from the $x$-axis. We define the normal vector ${\bm N}$ by
\begin{equation}
 {\bm N} = \left[\begin{array}{cc}0 &-1 \\1 & 0\end{array}\right]\frac{\partial\gamma}{\partial s}
=\left[\begin{array}{c}-\sin\theta\\\cos\theta\end{array}\right],
\end{equation}
and introduce the Frenet frame 
\begin{equation}
 F=({\bm T},{\bm N}),\quad {\bm T}=\frac{\partial\gamma}{\partial s},
\end{equation}
which is the orthonormal basis attached to the curve. The Frenet equation is given by
\begin{equation}
 \frac{\partial}{\partial s}F = F \left[\begin{array}{cc} 0 
&-\kappa \\\kappa & 0\end{array}\right],\label{F_s}
\end{equation}
where the function $\kappa=\frac{\partial\theta}{\partial s}$ is the curvature of $\gamma$. The angle function $\theta$ is
also referred to as the potential function.  Let us consider the following isoperimetric motion in
time $t$:
\begin{equation}
 \frac{\partial}{\partial t}F = F \left[
\begin{array}{cc}
0 &\kappa_{ss}+\frac{\kappa^3}{2} 
\\
-\kappa_{ss}-\frac{\kappa^3}{2} &0
\end{array}
\right].\label{F_t}
\end{equation}
In terms of $\frac{\partial\gamma}{\partial s}$, (\ref{F_s}) and (\ref{F_t}) can be expressed as
\begin{eqnarray}
 \frac{\partial^2\gamma}{\partial s^2}&=&\left[\begin{array}{cc} 0 
&-\kappa \\\kappa & 0\end{array}\right]\frac{\partial\gamma}{\partial s},\label{gamma_x}\\
 \frac{\partial}{\partial t}\left(\frac{\partial\gamma}{\partial s}\right)
&=&\left[
\begin{array}{cc}
0 &\kappa_{ss}+\frac{\kappa^3}{2} 
\\
-\kappa_{ss}-\frac{\kappa^3}{2} &0
\end{array}
\right]\frac{\partial\gamma}{\partial s},
\label{gamma_t}
\end{eqnarray}
respectively. Then the compatibility condition of (\ref{F_s}) and (\ref{F_t}), or (\ref{gamma_x})
and (\ref{gamma_t}) yields the mKdV equation for
$\kappa=\kappa(s,t)$~\cite{Goldstein-Petrich,Lamb}
\begin{equation}
 \kappa_t +\frac{3}{2}\kappa^2\kappa_s + \kappa_{sss}=0, \label{eqn:mKdV}
\end{equation}
or the potential mKdV equation for $\theta=\theta(s,t)$:
\begin{equation}
 \theta_t + \frac{1}{2}(\theta_s)^3 + \theta_{sss}=0.\label{eqn:pmKdV}
\end{equation}
The mKdV equation can be viewed as the governing equation of the Lagrangian description for the
motion of the curves $\gamma$ in terms of the arc-length parameter $s$.  Let us consider the
Eulerian description of the same motion of the curves. To this end, we introduce the Eulerian
coordinates
\begin{equation}
 \gamma(s,t)=\left[\begin{array}{l}x(s,t)
	       \\v(s,t)\end{array}\right]
=\int_0^s \left[\begin{array}{l}\cos\theta(s',t)
	       \\\sin\theta(s',t)\end{array}\right]ds' + 
\left[\begin{array}{l}x_0 \\v_0\end{array}\right],\label{hodograph:mKdV-WKI}
\end{equation}
and change the independent variables $(s,t)$ to 
\begin{equation}
(x,t') =\left(\int_0^s \cos\theta(s',t)~ds'+x_0,\ t\right).\label{hodograph2:mKdV-WKI}
\end{equation}
For simplicity we write $t'$ as $t$ without causing confusion.
Let us write down the equation for $v$ in terms of $x$ and $t$. It can be easily shown that
\begin{eqnarray}
&&s(x,t)=\int \sqrt{1+v_x^2}~dx,\quad \kappa(x,t)=\frac{v_{xx}}{(1+v_x^2)^{\frac{3}{2}}},\\
&&{\bm N} = \frac{1}{\sqrt{1+v_x^2}}\left[\begin{array}{c}-v_x\\1\end{array}\right],\quad
{\bm T} = \frac{1}{\sqrt{1+v_x^2}}\left[\begin{array}{c}1\\v_x\end{array}\right].
\end{eqnarray}
Noticing 
\begin{equation}
\frac{\partial}{\partial t}\gamma=-\kappa_s{\bf N} - \frac{1}{2}\kappa^2 {\bf T},\label{gamma_t:NT}
\end{equation}
it follows that
\begin{equation}
-\kappa_s=\gamma_t\cdot {\bf N}=\frac{v_t}{\sqrt{1+v_x^2}},
\end{equation}
by taking the inner product with ${\bm N}$ on both sides of (\ref{gamma_t:NT}). By using $\frac{ds}{dx}=\sqrt{1+v_x^2}$,
we see that 
\begin{equation}
 v_t=-\kappa_s {\sqrt{1+v_x^2}}=-\kappa_x.
\end{equation}
Thus we derive
\begin{equation}
v_t=-\left(\frac{v_{xx}}{(1+v_x^2)^{\frac{3}{2}}}\right)_x.\label{eqn:pWKI}
\end{equation}
Introducing $u=v_x$, we obtain the WKI (Wadati-Konno-Ichikawa) 
elastic beam equation~\cite{WKI1,WKI2,WKI3,Konno}
\begin{equation}
u_{t}=-\left(\frac{u_{x}}{(1+u^2)^{\frac{3}{2}}}\right)_{xx}  .\label{eqn:WKI}
\end{equation}
Therefore, (\ref{eqn:pWKI}) or (\ref{eqn:WKI}) can be viewed as the governing equation of the
Eulerian description for the curve motions given by (\ref{F_s}) and (\ref{F_t}).

We note that (\ref{hodograph:mKdV-WKI}) is the hodograph transformation between the potential mKdV
equation (\ref{eqn:pmKdV}) and the WKI elastic beam equation (\ref{eqn:pWKI}) found by
Ishimori~\cite{Ishimori}. The above discussion shows that the hodograph transformation arises
naturally as the transformation between the Lagrangian and Eulerian descriptions from the point of
view of geometry of plane curves.

There is another equation related to the plane curve motions which is
known as the complex Dym
equation (this is often called the complex Harry Dym equation)
~\cite{Goldstein-Petrich,WKI1,Kruskal,Kadanoff,Kawamoto,Dmitrieva,Hereman,WIS}.
Introducing the complex variables $r$ and $z$, we consider the transformation
\begin{equation}
 r(s,t) = e^{\sqrt{-1}~\theta(s,t)},
\end{equation}
\begin{equation}
 z(s,t) = \int_0^se^{\sqrt{-1}~\theta(s',t)~}ds'+z_0,\quad t'=t.
\end{equation}
Then the potential mKdV equation (\ref{eqn:pmKdV}) is transformed to 
the complex Dym equation~\cite{Kawamoto,Dmitrieva}
\begin{equation}
 r_t = r^3r_{zzz}.\label{complexDym}
\end{equation}
Here we set $t'=t$ without causing confusion. The geometric meaning of the complex Dym equation may
be described as follows. The variables $r$ and $z$ are expressed as
\begin{equation}
 r=\cos\theta(s,t) + \sqrt{-1}\sin\theta(s,t) = x_s + \sqrt{-1}v_s,\quad
z= x+\sqrt{-1}v+z_0.\label{r_and_z}
\end{equation}
Identifying the Euclidean plane $\mathbb{R}^2$ as $\mathbb{C}$, we see that $z$ is the position
vector of the curve $\gamma$, and $r$ corresponds to the tangent vector
$\frac{\partial\gamma}{\partial s}$. By treating the
Eulerian coordinates of $\gamma$ as a complex variable $z$, the complex Dym equation is nothing but
the governing equation to describe the motion of the tangent vector $\frac{\partial\gamma}{\partial s}$.

It is well known that the sine-Gordon equation
\begin{equation}
 \theta_{ys}=4\sin\theta, \label{eqn:sG}
\end{equation}
belongs to the the same hierarchy as the mKdV equation~\cite{AKNS,AS:book} and that it describes a
certain motion of plane curves~\cite{NSW:1992}. It is possible to derive the governing equation of
curve motion in the 
Eulerian description in a similar manner 
to the case of the mKdV equation. In fact,
applying the transformations
\begin{equation}
(x,y') =\left(\int_0^s \cos\theta(s',y)~ds'+x_0,\ y\right),
\label{hodograph:sG-sp}
\end{equation}
\begin{equation}
 v = \int_0^s \sin\theta(s',y)~ds'+v_0,
\end{equation}
we obtain the short pulse equation~\cite{SP,Rabelo,Beals,Brunelli1,Brunelli2}
\begin{equation}
 v_{xy} = 4v + \frac{2}{3}\left(v^3\right)_{xx}, \label{eqn:short-pulse}
\end{equation}
where we set $y'=y$ for simplicity. Again, we note that the short pulse equation
(\ref{eqn:short-pulse}) describes the same curve motions as the sine-Gordon equation by using the
Eulerian description. The transformation (\ref{hodograph:sG-sp}) gives the hodograph transformation
between them~\cite{Matsuno-SP,Sakovich,Sakovich2,Rogers}.
%
\subsection{Continuous motion of discrete curves}
In this subsection we discuss the semi-discrete equations arising from
the continuous motion of discrete plane curves and the hodograph transformations among them.
A map $\gamma:~\mathbb{Z}\rightarrow \mathbb{R}^2;~l \mapsto \gamma_l$ 
is said to be a discrete
curve of segment length $a_l$ if
\begin{equation}
 \left|\frac{\gamma_{l+1} - \gamma_l}{a_l}\right| = 1.
\label{iso0}
\end{equation}
We introduce the angle function $\psi_l$ of 
a discrete curve $\gamma$ by
\begin{equation}
 \frac{\gamma_{l+1} - \gamma_l}{a_l}=
\left[\begin{array}{c}\cos\psi_l \\\sin\psi_l \end{array}\right].
\label{discrete_angle}
\end{equation}
A discrete curve $\gamma$ satisfies
\begin{equation}
 \frac{\gamma_{l+1} - \gamma_{l}}{a_l} 
= R(\kappa_l)~\frac{\gamma_{l} - \gamma_{l-1}}{a_{l-1}},\label{discrete_FS}
\end{equation}
for $\kappa_l=\psi_{l}-\psi_{l-1}$, where $R(\kappa_l)$ 
denotes the rotation matrix given by
\begin{equation}
 R(\kappa_l)=\left(\begin{array}{cc}\cos \kappa_l & -\sin \kappa_l\\
\sin \kappa_l & \cos \kappa_l\end{array}\right).
\end{equation}
We set $a_l=\epsilon\,(>0)$, 
and consider the following motion of discrete curves:
\begin{equation}
 \frac{d\gamma_l}{d\zeta}
=\frac{1}{\cos\frac{\kappa_l}{2}}~R\left(-\frac{\kappa_l}{2}\right)~
\frac{\gamma_{l+1}-\gamma_{l}}{\epsilon}.\label{gamma_w}
\end{equation}
Then from the isoperimetric condition (\ref{iso0}) 
and the compatibility condition of (\ref{discrete_FS}) and (\ref{gamma_w})
, it follows that there exists a potential function $\theta_l$ 
characterized by
\begin{equation}
 \psi_l = \frac{\theta_{l+1}+\theta_l}{2},\quad \kappa_l =  
\frac{\theta_{l+1}-\theta_{l-1}}{2},\label{psi_and_theta}
\end{equation}
and that $\theta_l$ satisfies the semi-discrete potential 
mKdV equation~\cite{Ablowitz-Ladik:1975,DS1,DS2,DS3,Hirota:semi-discrete_mKdV,Hoffmann:LN,HK}
\begin{equation}
\frac{d\theta_l}{d\zeta} 
= \frac{2}{\epsilon}\tan\left(\frac{\theta_{l+1}-\theta_{l-1}}{4}\right). \label{sdpmKdV}
\end{equation}
We note that $K_l = \frac{2}{\epsilon}\tan\frac{\kappa_l}{2}$ satisfies the
semi-discrete mKdV equation
\begin{equation}
\frac{dK_l}{d\zeta} = \frac{2}{\epsilon}
\left(1+\frac{\epsilon^2}{4}K_l^2\right)(K_{l+1}-K_{l-1}).\label{sdmKdV}
\end{equation}

It is possible to consider the Eulerian description of the curve motion defined by
(\ref{discrete_FS}) and (\ref{gamma_w}). 
Noticing (\ref{discrete_angle}) and (\ref{psi_and_theta}),
we introduce the Eulerian coordinates
\begin{equation}
 \gamma_l(\zeta) = \left[\begin{array}{c} X_l(\zeta)\\v_l(\zeta)\end{array}\right]=
\sum_{j=0}^{l-1} 
\left[\begin{array}{c}\medskip\epsilon\cos\left(\frac{\theta_{j+1}+\theta_j}{2}\right) \\
\epsilon\sin\left(\frac{\theta_{j+1}+\theta_j}{2}\right)  \end{array} \right]
+\left[\begin{array}{c} X_0\\v_0\end{array}\right].\label{semi-discrete_Euler}
\end{equation}
 
Then from (\ref{discrete_angle}), (\ref{sdpmKdV}) 
and (\ref{semi-discrete_Euler}), one can derive
\begin{eqnarray}
&&\frac{d}{d\zeta}\delta_l
=-\frac{v_{l+1}-v_l}{\epsilon}
\left(
\frac{\Delta_{l+1} - \Delta_l}{1 + \Delta_{l+1}\Delta_{l}}
+
\frac{\Delta_l - \Delta_{l-1}}{1 + \Delta_l \Delta_{l-1}} \right), \label{eqn:sdWKI1}\\[2mm]
&&\frac{d}{d\zeta}\left(v_{l+1}-v_l\right)
=\frac{\delta_l}{\epsilon}
\left(
\frac{\Delta_{l+1} - \Delta_l}{1 + \Delta_{l+1}\Delta_{l}}
+
\frac{\Delta_l - \Delta_{l-1}}{1 + \Delta_l \Delta_{l-1}} \right),\label{eqn:sdWKI2}
\end{eqnarray}
where
\begin{equation}
\delta_l = X_{l+1}-X_{l},\quad \Delta_l 
= {\rm sgn}~(v_{l+1}-v_l)\left|\frac{v_{l+1}-v_{l}}{\epsilon+\delta_{l}}
\right|=\frac{v_{l+1}-v_{l}}{\epsilon+\delta_{l}}
, \label{eqn:sdWKI3}
\end{equation}
since $\epsilon+\delta_{l}=\epsilon(1+\cos \psi_l)\geq 0$.
Note that $v_l$ and $\delta_l$ satisfy
\begin{equation}
\left(\frac{v_{l+1}-v_l}{\epsilon}\right)^2 
+ \left(\frac{\delta_l}{\epsilon}\right)^2=1. 
\end{equation}
From (\ref{eqn:sdWKI1}) and (\ref{eqn:sdWKI2}), we obtain 
\begin{equation}
\frac{d}{d\zeta}\left(\frac{v_{l+1}-v_l}{X_{l+1}-X_l}\right)
=\frac{1}{\epsilon}\left(1+\left(\frac{v_{l+1}-v_l}
{X_{l+1}-X_l}\right)^2\right)
\left(
\frac{\Delta_{l+1} - \Delta_l}{1 + \Delta_{l+1}\Delta_{l}}
+
\frac{\Delta_l - \Delta_{l-1}}{1 + \Delta_l \Delta_{l-1}} \right)\,.\label{eqn:sdWKI4}
\end{equation}
The system of (\ref{eqn:sdWKI1}), (\ref{eqn:sdWKI2}) and
(\ref{eqn:sdWKI3}) 
is nothing but the
semi-discrete WKI elastic beam equation. We remark that (\ref{semi-discrete_Euler}) can be regarded
as the hodograph transformation between (\ref{sdpmKdV}) and the semi-discrete WKI elastic beam
equation. Note that the angle function $\psi_l=\frac{\theta_{l+1}+\theta_l}{2}$ satisfies
\begin{equation}
\cos \psi_l=\frac{X_{l+1}-X_l}{\epsilon}\,,\qquad 
\sin \psi_l=\frac{v_{l+1}-v_l}{\epsilon}\,,\qquad 
\tan \psi_l=\frac{v_{l+1}-v_l}{X_{l+1}-X_l}\,.
\end{equation}
Thus (\ref{eqn:sdWKI4}) can be rewritten as 
\begin{equation}
\frac{d}{d\zeta}\psi_l=\frac{1}{\epsilon}\left(
\tan \frac{\psi_{l+1}-\psi_{l}}{2}+
\tan \frac{\psi_{l}-\psi_{l-1}}{2}
\right)\,.\label{eqn:sdWKI5}
\end{equation}
Equation (\ref{eqn:sdWKI5}) with the discrete hodograph transformation
\begin{equation}
X_l(\zeta)=
\sum_{j=0}^{l-1} \epsilon\cos \psi_j(\zeta)+X_0\,,\quad 
v_l(\zeta)=\sum_{j=0}^{l-1} 
\epsilon\sin \psi_l(\zeta)+v_0\,,\label{eqn:sdWKI6}
\end{equation}
can be also regarded as the semi-discrete WKI elastic beam equation. 
In the continuous limit $\epsilon \to 0$ with $s=\epsilon l+\zeta$ and 
$t=-\frac{\epsilon^2}{6}\zeta$, 
(\ref{eqn:sdWKI5}) and (\ref{eqn:sdWKI6}) converge to 
\begin{equation}
 \theta_t + \frac{1}{2}(\theta_s)^3 + \theta_{sss}=0\,,
\end{equation}
and 
\begin{equation}
x(s,t)=\int_0^{s}\cos \theta(s',t) ds'+x_0\,,\quad 
v(s,t)=\int_0^{s}\sin \theta(s',t) ds'+v_0\,,
\end{equation}
which give the (potential) WKI elastic beam equation (\ref{eqn:pWKI}) 
(see Appendix). 

%
One can construct a semi-discrete version of the complex 
Dym equation as follows. 
In view of (\ref{r_and_z}), (\ref{psi_and_theta}) and (\ref{semi-discrete_Euler}), it is natural to 
introduce the complex variables $r_l(\zeta)$ and $Z_l(\zeta)$ by
\begin{equation}
\fl
 r_l(\zeta)= e^{\sqrt{-1}\psi_l}
 = e^{\sqrt{-1}\frac{\theta_{l+1} + \theta_l}{2}},\quad Z_l(\zeta) = X_l(\zeta)+\sqrt{-1}v_l(\zeta)=\sum_{j=0}^{l-1} \epsilon
  e^{\sqrt{-1}~\frac{\theta_{j+1}+\theta_j}{2}} + Z_0.
\label{hodograph:semi-discrete_Dym}
\end{equation}
Then we have from (\ref{sdpmKdV}) and (\ref{hodograph:semi-discrete_Dym})
\begin{equation}
 \frac{dr_l}{d\zeta} =
  \frac{r_l}{\epsilon}\left(\frac{r_{l+1}-r_l}{r_{l+1}+r_l}
+\frac{r_{l}-r_{l-1}}{r_{l}+r_{l-1}}\right),\quad
  \frac{Z_{l+1}-Z_{l}}{\epsilon} = r_l,\label{semi-discrete_Dym}
\end{equation}
which is the semi-discrete complex Dym equation. 
The geometric meaning of (\ref{semi-discrete_Dym})
can be described as follows: under the identification of $\mathbb{R}^2$ as
$\mathbb{C}$, $Z_l$ is the position vector of the curve $\gamma_l$, 
and $r_l$ corresponds to
the segment vector 
$\frac{\gamma_{l+1}-\gamma_l}{\epsilon}$. Then (\ref{semi-discrete_Dym}) 
is the governing equation
describing the motion of the segment vector in the Eulerian coordinates 
of $\gamma_l$. To take the continuous limit, we use the angle function
$\psi_l$. Then the semi-discrete
complex Dym equation (\ref{semi-discrete_Dym}) is rewritten as 
\begin{equation}
\frac{d}{d\zeta}\psi_l=\frac{1}{\epsilon}\left(
\tan \frac{\psi_{l+1}-\psi_{l}}{2}+
\tan \frac{\psi_{l}-\psi_{l-1}}{2}
\right)\,.\label{eqn:sdDym1}
\end{equation}
and
\begin{equation}
Z_l(\zeta)=
\sum_{j=0}^{l-1} \epsilon e^{\sqrt{-1}\psi_j(\zeta)}+Z_0\,,\quad 
r_l(\zeta)=e^{\sqrt{-1}\psi_l(\zeta)}\,.\label{eqn:sdDym2}
\end{equation}
In the continuous limit $\epsilon \to 0$ with $s=\epsilon l+\zeta$ and 
$t=-\frac{\epsilon^2}{6}\zeta$, 
(\ref{eqn:sdDym1}) and (\ref{eqn:sdDym2}) converge to 
\begin{equation}
 \theta_t + \frac{1}{2}(\theta_s)^3 + \theta_{sss}=0\,,
\end{equation}
and 
\begin{equation}
z(s,t)=\int_0^{s}e^{\sqrt{-1}\theta(s',t)} ds'+z_0\,,\quad 
r(s,t)=e^{\sqrt{-1}\theta(s,t)}\,,
\end{equation}
which give the complex Dym equation (\ref{complexDym}) (see Appendix). 

%
Now we construct the semi-discrete short pulse equation. To this end, we consider the semi-discrete
sine-Gordon equation
\begin{equation}
 \frac{d}{dy}(\theta_{l+1}-\theta_l) = 4\epsilon
  \sin\left(\frac{\theta_{l+1}+\theta_{l-1}}{2}\right).\label{eqn:semi-discrete_sG}
\end{equation}
Similar to the continuous case, the semi-discrete sine-Gordon equation (\ref{eqn:semi-discrete_sG})
can be regarded as describing a certain motion of discrete plane curves.  Therefore, we may expect
that the application of the same transformation as the case of the
semi-discrete WKI equation to the semi-discrete sine-Gordon equation 
(\ref{eqn:semi-discrete_sG}) yields the semi-discrete analogue of the short pulse equation. In
fact, by using the transformation
\begin{equation}
 \gamma_l(y) = \left[\begin{array}{c} X_l(y)\\v_l(y)\end{array}\right]=
\sum_{j=0}^{l-1} 
\left[\begin{array}{c}\medskip\epsilon\cos\left(\frac{\theta_{j+1}+\theta_j}{2}\right) \\
\epsilon\sin\left(\frac{\theta_{j+1}+\theta_j}{2}\right)  \end{array} \right]
+\left[\begin{array}{c} X_0\\v_0\end{array}\right],\label{semi-discrete_Euler2}
\end{equation}
we obtain the semi-discrete short pulse equation
\begin{eqnarray}
 &&\frac{d}{dy}(X_{l+1}-X_l) = -2(v_{l+1}^2-v_l^2), \label{semi-discrete_SP1}\\
&&\frac{d}{dy}(v_{l+1}-v_l) = 2(X_{l+1}-X_l)(v_{l+1}+v_l). \label{semi-discrete_SP2}
\end{eqnarray}
We note that the following relation also holds from (\ref{semi-discrete_Euler2})
\begin{equation}
 \left(\frac{v_{l+1}-v_l}{\epsilon}\right)^2 +  \left(\frac{X_{l+1}-X_l}{\epsilon}\right)^2=1.
\end{equation}
From (\ref{semi-discrete_SP1}) and (\ref{semi-discrete_SP2}), we obtain 
\begin{equation}
\frac{d}{dy}\left(\frac{v_{l+1}-v_l}{X_{l+1}-X_l}\right)
=2(v_{l+1}+v_{l})+2\left(\frac{v_{l+1}-v_l}{X_{l+1}-X_l}\right)^2
(v_{l+1}+v_{l})\,.\label{semi-discrete_SP3}
\end{equation}
In order to take the continuous limit, we assume the boundary condition
$X_l=v_l=0$ for $l<0$, which is consistent with (\ref{semi-discrete_Euler2}).
Then the continuous limit $\epsilon\to 0$ (i.e., $X_{l+1}-X_l \to 0$) gives
\begin{eqnarray*}
&&\frac{v_{l+1}-v_l}{X_{l+1}-X_l}\to \frac{\partial v}{\partial x}\,,
\quad \frac{v_{l+1}+v_l}{2}\to v\,,\\
&&\frac{\partial X_l}{\partial y}=\frac{\partial X_0}{\partial y}
+\sum_{j=0}^{l-1}\frac{\partial (X_{j+1}-X_j)}{\partial y} 
=\frac{\partial X_0}{\partial y}
-2\sum_{j=0}^{l-1}(v_{j+1}^2-v_j^2)
=-2v_{l}^2\to \frac{\partial x}{\partial y}= -2 v^2\,,\\
&& \frac{\partial}{\partial y}=\frac{\partial}{\partial
 y'}+\frac{\partial x}{\partial y}\frac{\partial }{\partial x}
= \frac{\partial}{\partial
 y'}-2v_l^2\frac{\partial }{\partial x} \to 
\frac{\partial}{\partial y'}-2v^2\frac{\partial }{\partial x}\,. 
\end{eqnarray*}
Consequently, (\ref{semi-discrete_SP3}) converges to 
\begin{equation}
(\partial_{y'}-2v^2\partial_x)v_x=4v+4vv_x^2\,,
\end{equation}
which is nothing but the short pulse equation (\ref{eqn:short-pulse}).
\par\bigskip

%
\subsection{Discrete motion of discrete curves}


Now let us recall the following discrete motion of discrete plane curve
$\gamma_n^m$ introduced 
by Matsuura~\cite{Matsuura}:
\begin{eqnarray}
& \left|\frac{\gamma^{m}_{n+1} - \gamma^{m}_n}{a_n}\right| = 1,
 \label{iso:full}\\[2mm]
& \frac{\gamma^{m}_{n+1} - \gamma^{m}_{n}}{a_n} = R(\kappa^m_n)~\frac{\gamma^{m}_{n} - \gamma^{m}_{n-1}}{a_{n-1}},\label{gamma_n}\\[2mm]
& \frac{\gamma^{m+1}_{n} - \gamma^{m}_{n}}{b_m} = R(\omega^m_n)~\frac{\gamma^{m}_{n+1} - \gamma^{m}_{n}}{a_{n}},
\label{gamma_m}
\end{eqnarray}
where $a_n$ and $b_m$ are arbitrary functions in $n$ and $m$,
respectively. Compatibility of the system (\ref{iso:full})--(\ref{gamma_m}) 
implies the existence of
the potential function $\theta_n^m$ defined by
\begin{equation}
\omega^n_m =\frac{\theta^{m+1}_{n} - \theta^{m}_{n+1}}{2}, \quad
\kappa^m_n = \frac{\theta^{m}_{n+1} - \theta^{m}_{n-1}}{2},
\end{equation}
and it follows that $\theta_n^m$ satisfies the 
discrete potential mKdV equation~\cite{Hirota:dpmKdV}:
\begin{equation}
  \tan\left(\frac{\theta_{n+1}^{m+1}-\theta_{n}^m}{4}\right) 
= \frac{b_m+a_n}{b_m-a_n}~\tan\left(\frac{\theta_{n}^{m+1} 
- \theta_{n+1}^{m}}{4}\right).\label{eqn:dpmKdV} 
\end{equation}
Note that the functions $\psi_n^m$ and $\phi_n^m$ can be expressed as
\begin{equation}
 \psi_n^m = 
\frac{\theta_{n+1}^m+\theta_n^m}{2}\,,\quad 
\phi_n^m = \frac{\theta_{n}^{m+1}+\theta_n^m}{2}\,.
\label{discrete_angle_and_potential}
\end{equation}
Note also
\begin{eqnarray}
\frac{\gamma_{n+1}^m-\gamma_n^m}{a_n}=
\left[
\begin{array}{c}
\cos \psi_n^m\\
\sin \psi_n^m
\end{array}
\right]\,,\quad 
\frac{\gamma_{n}^{m+1}-\gamma_n^m}{b_m}=
\left[
\begin{array}{c}
\cos \phi_n^m\\
\sin \phi_n^m
\end{array}
\right]\,,
\end{eqnarray}
and
\begin{equation}
\gamma_n^m=
\left[
\begin{array}{c}
X_n^m\\
v_n^m
\end{array}
\right]
=
\sum_{j=0}^{n-1}
\left[
\begin{array}{c}
X_{j+1}^m-X_j^m\\
v_{j+1}^m-v_j^m
\end{array}
\right]
+
\left[
\begin{array}{c}
X_0^m\\
v_0^m
\end{array}
\right]
=\sum_{j=0}^{n-1}
a_j\left[
\begin{array}{c}
\cos \psi_j^m\\
\sin \psi_j^m
\end{array}
\right]
+
\left[
\begin{array}{c}
X_0^m\\
v_0^m
\end{array}
\right]\,.\label{hodograph:discrete}
\end{equation}
From the discrete potential mKdV equation (\ref{eqn:dpmKdV}) 
and the hodograph
transformation (\ref{hodograph:discrete}), we obtain 
\begin{equation}
\frac{\Delta_n^{m+1}-\Gamma_n^m}{1+\Delta_n^{m+1}\Gamma_n^m}
=\frac{b_m+a_n}{b_m-a_n}
\frac{\Gamma_n^m-\Delta_n^m}{1+\Gamma_n^m\Delta_n^m}\,,\label{discreteWKI}
\end{equation}
where 
\begin{eqnarray}
&&\Delta_n^m={\rm sgn}(v_{n+1}^m-v_n^m)
\left|\frac{v_{n+1}^m-v_n^m}{a_n+(X_{n+1}^m-X_n^m)}\right|
=\frac{v_{n+1}^m-v_n^m}{a_n+(X_{n+1}^m-X_n^m)}
\,,\label{discreteWKI-1}\\
&&\Gamma_n^m={\rm sgn}(v_{n}^{m+1}-v_n^m)
\left|\frac{v_{n}^{m+1}-v_n^m}{b_m+(X_{n}^{m+1}-X_n^m)}\right|
=\frac{v_{n}^{m+1}-v_n^m}{b_m+(X_{n}^{m+1}-X_n^m)}
\,,\label{discreteWKI-2}
\end{eqnarray}
since $a_n+(X_{n+1}^m-X_n^m)=a_n(1+\cos \psi_n^m)\geq 0$ and 
$b_m+(X_{n}^{m+1}-X_n^m)=b_m(1+\cos \phi_n^m)\geq 0$.
We note that $v_n^m$ and $X_n^m$ satisfy the following relations
\begin{eqnarray}
&&\left(\frac{X_{n+1}^m-X_n^m}{a_n}\right)^2+
\left(\frac{v_{n+1}^m-v_n^m}{a_n}\right)^2=1\,,\label{constraint1}\\
&&\left(\frac{X_{n}^{m+1}-X_n^m}{b_m}\right)^2+
\left(\frac{v_{n}^{m+1}-v_n^m}{b_m}\right)^2=1\,.\label{constraint2}
\end{eqnarray}

To construct an explicit form of the discrete WKI elastic beam equation, 
we consider an identity
\begin{equation}
e^{\sqrt{-1}\psi_n^{m+1}}e^{\sqrt{-1}\psi_n^{m}}
=e^{\sqrt{-1}\phi_{n+1}^{m}}e^{\sqrt{-1}\phi_n^{m}}\,.\label{identity:wki}
\end{equation}
Substituting
\begin{eqnarray*}
e^{\sqrt{-1}\psi_n^{m}}=\cos\psi_n^m+\sqrt{-1}\sin \psi_n^m=\frac{X_{n+1}^m-X_n^m}{a_n}+\sqrt{-1}~\frac{v_{n+1}^m-v_n^m}{a_n}\,,\\
e^{\sqrt{-1}\phi_n^{m}}=\cos\phi_n^m+\sqrt{-1}\sin \phi_n^m=
\frac{X_{n}^{m+1}-X_n^m}{b_m}
+\sqrt{-1}~\frac{v_{n}^{m+1}-v_n^m}{b_m}\,,
\end{eqnarray*}
into (\ref{identity:wki}), we obtain the system of two discrete equations from the real and
imaginary parts, respectively, which should be considered together with the constraints (\ref{constraint1}) and
(\ref{constraint2}). To incorporate (\ref{constraint1}) and (\ref{constraint2}), an easy way is to
consider the identity
\begin{equation}
e^{\sqrt{-1}\psi_n^{m+1}}e^{-\sqrt{-1}\phi_n^{m}}
=e^{\sqrt{-1}\phi_{n+1}^{m}}e^{-\sqrt{-1}\psi_n^{m}}\,,\label{identity2:wki}
\end{equation}
instead of (\ref{identity:wki}). We then obtain the following system of two discrete equations
\begin{eqnarray}
&&(X_{n+1}^{m+1}-X_{n}^{m+1})(X_{n}^{m+1}-X_{n}^{m})
+(v_{n+1}^{m+1}-v_{n}^{m+1})(v_{n}^{m+1}-v_{n}^{m})\nonumber\\
&&\qquad =(X_{n+1}^{m+1}-X_{n+1}^{m})(X_{n+1}^{m}-X_{n}^{m})
+(v_{n+1}^{m+1}-v_{n+1}^{m})(v_{n+1}^{m}-v_{n}^{m})\,,\label{discreteWKI1}\\
&&(X_{n+1}^{m+1}-X_{n}^{m+1})(v_{n}^{m+1}-v_{n}^{m})
-(X_{n}^{m+1}-X_{n}^{m})(v_{n+1}^{m+1}-v_{n}^{m+1})\nonumber\\
&&\qquad =(X_{n+1}^{m+1}-X_{n+1}^{m})(v_{n+1}^{m}-v_{n}^{m})
-(X_{n+1}^{m}-X_{n}^{m})(v_{n+1}^{m+1}-v_{n+1}^{m})\,,\label{discreteWKI2}
\end{eqnarray}
which is simplified to 
\begin{eqnarray}
&&(X_{n+1}^{m+1}-X_{n+1}^{m}-X_{n}^{m+1}+X_{n}^{m})
(X_{n+1}^{m}-X_{n}^{m+1})\nonumber\\
&&\qquad +
(v_{n+1}^{m+1}-v_{n+1}^{m}-v_{n}^{m+1}+v_{n}^{m})
(v_{n+1}^{m}-v_{n}^{m+1})=0\,,\label{eqn:discreteWKI1}\\
&&
(X_{n+1}^{m+1}+X_{n+1}^{m}-X_{n}^{m+1}-X_{n}^{m})
(v_{n+1}^{m+1}-v_{n+1}^{m}+v_{n}^{m+1}-v_{n}^{m})\nonumber\\
&&\qquad -(X_{n+1}^{m+1}-X_{n+1}^{m}+X_{n}^{m+1}-X_{n}^{m})
(v_{n+1}^{m+1}+v_{n+1}^{m}-v_{n}^{m+1}-v_{n}^{m})=0\,.\label{eqn:discreteWKI2}
\end{eqnarray}
Note that the second equation (\ref{eqn:discreteWKI2}) is further
simplified to 
\begin{equation}
(X_{n+1}^{m}-X_{n}^{m+1})(v_{n+1}^{m+1}-v_{n}^{m})
-(X_{n+1}^{m+1}-X_{n}^{m})(v_{n+1}^{m}-v_{n}^{m+1})=0\,.\label{eqn:discreteWKI3}
\end{equation}
We remark that the system (\ref{eqn:discreteWKI1}) and (\ref{eqn:discreteWKI2}) can be solved
explicitly in terms of $v_{n+1}^{m+1}$ and $X_{n+1}^{m+1}$ in the form of rational functions 
of $v_n^m$, $v_{n+1}^{m}$, $v_{n+1}^{m+1}$, $X_{n}^{m}$, $X_{n+1}^{m}$ and $X_{n}^{m+1}$. Therefore
the system (\ref{eqn:discreteWKI1}) and (\ref{eqn:discreteWKI2}) (or (\ref{eqn:discreteWKI3})) can
be regarded as the explicit form of the discrete WKI elastic beam
equation. Note that we can obtain
(\ref{discreteWKI}), (\ref{discreteWKI-1}) and (\ref{discreteWKI-2}) 
by replacing $X_n^m$ by 
$X_n^m=X_n^m+\sum_{j=0}^{n-1}a_j+\sum_{j=0}^{m-1}b_j$ in
(\ref{discreteWKI1}) and (\ref{discreteWKI2}) and then dividing
(\ref{discreteWKI2}) by (\ref{discreteWKI1}). 
By using the potential function $\theta_n^m$, 
the discrete WKI elastic beam equation can be written as 
\begin{eqnarray}
&& \tan\left(\frac{\theta_{n+1}^{m+1}-\theta_{n}^m}{4}\right) 
= \frac{b_m+a_n}{b_m-a_n}~\tan\left(\frac{\theta_{n}^{m+1} 
- \theta_{n+1}^{m}}{4}\right)\,,\label{discreteWKI1-cont}\\
&&X_n^m=\sum_{j=0}^{n-1}
a_j\cos\left(\frac{\theta_{j+1}^m+\theta_{j}^m}{2}\right)+X_0^m\,. 
\label{discreteWKI2-cont}
\end{eqnarray}
Setting
\begin{equation}
\fl \zeta=(n+m)\delta\,,\quad l=n-m\,,\quad 
a_n=a\,,\quad b_m=b\,,\quad \delta=\frac{a+b}{2}\,,\quad 
\epsilon=\frac{a-b}{2}\,,
\end{equation}
and taking the continuous limit $\delta \to 0$, (\ref{discreteWKI1-cont}) and
(\ref{discreteWKI2-cont}) are reduced to
\begin{eqnarray}
&& 
\frac{d\theta_l}{d\zeta} 
= \frac{2}{\epsilon}\tan\left(\frac{\theta_{l+1}-\theta_{l-1}}{4}\right)
\,,\label{semi-discreteWKI1-cont}\\
&&X_l(\zeta)=\sum_{j=0}^{l-1}
a_j\cos\left(\frac{\theta_{j+1}(\zeta)+\theta_{j}(\zeta)}{2}\right)
+X_0(\zeta)\,, 
\label{semi-discreteWKI2-cont}
\end{eqnarray}
which are transformed to the semi-discrete WKI elastic beam equation 
(\ref{eqn:sdWKI1}), (\ref{eqn:sdWKI2}) and
(\ref{eqn:sdWKI3}). 

Let us consider a discrete analogue of the complex Dym equation. 
Introducing 
\begin{eqnarray}
&&r_n^m=e^{\sqrt{-1}\frac{\theta_{n+1}^m+\theta_n^m}{2}}\,,\\
&&Z_n^m=X_n^m+\sqrt{-1}v_n^m=
\sum_{j=0}^{n-1}
a_j\cos \psi_j^m
+\sqrt{-1}\sum_{j=0}^{n-1}
a_j\sin \psi_j^m+Z_0^m\nonumber\\
&&\quad =
\sum_{j=0}^{n-1}a_jr_j^m+Z_0^m\,,
\end{eqnarray}
and using the discrete potential mKdV equation (\ref{eqn:dpmKdV}), 
we derive the discrete analogue of the complex Dym equation
\begin{eqnarray}
&& \frac{r_n^{m+1}-\rho_n^m}{r_n^{m+1}+\rho_n^m}=\frac{b_m+a_n}{b_m-a_n}~\frac{\rho_n^{m}-r_n^m}{\rho_n^{m}+r_n^m},\quad
\frac{r_n^{m+1}}{\rho_n^m} = \frac{\rho_{n+1}^m}{r_n^m},\label{discrete_Dym1}\\[2mm]
&&Z_{n+1}^m-Z_{n}^m=a_nr_n^m\,,\label{discrete_Dym2}
\end{eqnarray}
where $\rho_n^m$ is an auxiliary variable defined by
\begin{equation}
 \rho_n^m = e^{\sqrt{-1}\frac{\theta_{n}^{m+1}+\theta_n^m}{2}}\,.
\end{equation}
Similar to the semi-discrete case, the system of 
(\ref{discrete_Dym1}) and (\ref{discrete_Dym2}) describes the
motion of segment vector of the curve $\gamma_n^m$ in the Eulerian coordinates of $\gamma_n^m$.
It should be noted that by introducing $Q_n^m$ by
\begin{equation}
 Q_n^m=e^{\sqrt{-1}\theta_n^m}\,,
\end{equation}
we have an alternate form of the discrete complex Dym equation
\begin{eqnarray}
&&r_n^m=\sqrt{Q_{n+1}^mQ_n^m}\,,\label{discrete_Dym3}\\
&&\frac{\sqrt{Q_{n+1}^{m+1}}-\sqrt{Q_{n}^{m}}}
{\sqrt{Q_{n+1}^{m+1}}+\sqrt{Q_{n}^{m}}}
=\frac{b_m+a_n}{b_m-a_n}~
\frac{\sqrt{Q_{n}^{m+1}}-\sqrt{Q_{n+1}^{m}}}
{\sqrt{Q_{n}^{m+1}}+\sqrt{Q_{n+1}^{m}}}\,,\label{discrete_Dym4}\\
&&Z_{n+1}^m-Z_{n}^m=a_nr_n^m\,.\label{discrete_Dym5}
\end{eqnarray}
Using $\theta_n^m$, the discrete complex Dym equation can be written as 
\begin{eqnarray}
&& \tan\left(\frac{\theta_{n+1}^{m+1}-\theta_{n}^m}{4}\right) 
= \frac{b_m+a_n}{b_m-a_n}~\tan\left(\frac{\theta_{n}^{m+1} 
- \theta_{n+1}^{m}}{4}\right)\,,\label{discrete_Dym6}\\
&&Z_n^m=\sum_{j=0}^{n-1}
a_je^{\sqrt{-1}\frac{\theta_{j+1}^m+\theta_{j}^m}{2}}+Z_0^m\,. 
\label{discrete_Dym7}
\end{eqnarray}
Setting
\begin{equation}
\fl \zeta=(n+m)\delta\,,\quad l=n-m\,,\quad 
a_n=a\,,\quad b_m=b\,,\quad \delta=\frac{a+b}{2}\,,\quad 
\epsilon=\frac{a-b}{2}\,,
\end{equation}
and taking the continuous limit $\delta \to 0$, 
(\ref{discrete_Dym6}) and (\ref{discrete_Dym7}) become 
\begin{eqnarray}
&& 
\frac{d\theta_l}{d\zeta} 
= \frac{2}{\epsilon}\tan\left(\frac{\theta_{l+1}-\theta_{l-1}}{4}\right)
\,,\label{semi-discrete_Dym6}\\
&&Z_l(\zeta)=\sum_{j=0}^{l-1}
a_je^{\sqrt{-1}\frac{\theta_{j+1}(\zeta)+\theta_{j}(\zeta)}{2}}
+Z_0(\zeta)\,, 
\label{semi-discrete_Dym7}
\end{eqnarray}
which are transformed to the semi-discrete complex Dym equation 
(\ref{semi-discrete_Dym}).


We next construct the discrete short pulse equation. 
Consider the following discrete motion of plane 
discrete curve $\gamma_n^k$\cite{Inoguchi}
\begin{eqnarray}
& \left|\frac{\gamma^{k}_{n+1} - \gamma^{k}_n}{a_n}\right| = 1,\\
& \frac{\gamma^{k}_{n+1} - \gamma^{k}_{n}}{a_n} = R(\kappa^k_n)~
\frac{\gamma^{k}_{n} - \gamma^{k}_{n-1}}{a_{n-1}},\\
& \gamma^{k+1}_{n} - S\gamma^{k}_{n} = 
\frac{1}{c_k}SR(-\sigma^k_n)~\frac{\gamma^{k}_{n+1} -
\gamma^{k}_{n}}{a_{n}},\quad 
S=\left(\begin{array}{cc}1 &0 \\ 0 &-1 \end{array}\right)\,,
\end{eqnarray}
where $a_n$ and $c_k$ are arbitrary functions in $n$ and $k$,
respectively. 
Compatibility of this system implies the existence of
the potential function $\theta_n^k$ defined by
\begin{equation}
\kappa^k_n = \frac{\theta^{k}_{n+1} - \theta^{k}_{n-1}}{2},
\quad 
\sigma^k_n =\frac{\theta^{k+1}_{n} + \theta^{k}_{n+1}}{2}, 
\end{equation}
and it follows that $\theta_n^k$ satisfies the discrete 
sine-Gordon equation
\begin{equation}
\sin\left(\frac{\theta_{n+1}^{k+1}-\theta_{n+1}^{k}-
\theta_{n}^{k+1}+\theta_{n}^k}{4}\right) =
a_nc_k~\sin
\left(
\frac{\theta_{n+1}^{k+1}+\theta_{n+1}^{k}+
\theta_{n}^{k+1}+\theta_{n}^{k}}{4}\right)\,.\label{eqn:dsg}
\end{equation}
Note that the functions $\psi_n^k$ and $\varphi_n^k$ can be expressed as
\begin{equation}
 \psi_n^k = 
\frac{\theta_{n+1}^k+\theta_n^k}{2},\quad 
\varphi_n^k = \frac{\theta_{n}^{k+1}-\theta_n^k}{2}\,.
\label{discrete_angle_and_potential2}
\end{equation}
Note also
\begin{eqnarray}
\frac{\gamma_{n+1}^k-\gamma_n^k}{a_n}=
\left[
\begin{array}{c}
\cos \psi_n^k\\
\sin \psi_n^k
\end{array}
\right]\,,\quad 
\frac{\gamma_{n}^{k+1}-S\gamma_n^k}{\frac{1}{c_k}}=
\left[
\begin{array}{c}
\cos \varphi_n^k\\
\sin \varphi_n^k
\end{array}
\right]\,,
\end{eqnarray}
and
\begin{equation}
\gamma_n^k=
\left[
\begin{array}{c}
X_n^k\\
v_n^k
\end{array}
\right]
=
\sum_{j=0}^{n-1}
\left[
\begin{array}{c}
X_{j+1}^k-X_j^k\\
v_{j+1}^k-v_j^k
\end{array}
\right]
+
\left[
\begin{array}{c}
X_0^k\\
v_0^k
\end{array}
\right]
=\sum_{j=0}^{n-1}
a_j\left[
\begin{array}{c}
\cos \psi_j^k\\
\sin \psi_j^k
\end{array}
\right]
+
\left[
\begin{array}{c}
X_0^k\\
v_0^k
\end{array}
\right]\,.\label{hodograph:discrete:sp}
\end{equation}
From the discrete sine-Gordon equation (\ref{eqn:dsg}) 
and the hodograph
transformation (\ref{hodograph:discrete:sp}), we obtain 
\begin{equation}
\Delta_n^{k+1}-\Delta_n^{k}
=a_nc_k(\Gamma_n^{k+1}+\Gamma_n^k)\,,
\end{equation}
where 
\begin{eqnarray}
\Delta_n^k={\rm sgn}(v_{n+1}^k-v_n^k)
\left|\frac{v_{n+1}^k-v_n^k}{a_n+(X_{n+1}^k-X_n^k)}\right|
=\frac{v_{n+1}^k-v_n^k}{a_n+(X_{n+1}^k-X_n^k)}\,,
\\
\Gamma_n^k={\rm sgn}(v_{n}^{k+1}+v_n^k)
\left|\frac{v_{n}^{k+1}+v_n^k}{\frac{1}{c_k}
+(X_{n}^{k+1}-X_n^k)}\right|
=\frac{v_{n}^{k+1}+v_n^k}{\frac{1}{c_k}
+(X_{n}^{k+1}-X_n^k)}
\,,
\end{eqnarray}
since $a_n+(X_{n+1}^k-X_n^k)=a_n(1+\cos\psi_n^k)\geq 0$ and 
$\frac{1}{c_k}+(X_{n}^{k+1}-X_n^k)=\frac{1}{c_k}(1+\cos\varphi_n^k)\geq
0$. 
We note that $v_n^k$ and $X_n^k$ satisfy the following relations
\begin{eqnarray}
&&\left(\frac{v_{n+1}^k-v_n^k}{a_n}\right)^2
+\left(\frac{X_{n+1}^k-X_n^k}{a_n}\right)^2
=1\,,\label{constraint1:sp}\\
&&\left(\frac{v_n^{k+1}+v_n^k}{\frac{1}{c_k}}\right)^2
+\left(\frac{X_{n}^{k+1}-X_n^k}{\frac{1}{c_k}}\right)^2=
1
\,. \label{constraint2:sp}
\end{eqnarray}
We now construct an explicit form of the discrete short pulse equation.
Similar to the case of discrete WKI elastic beam equation, we consider
the identity
\begin{equation}
e^{\sqrt{-1}\psi_n^{k+1}}e^{-\sqrt{-1}\varphi_n^{k}}
=e^{\sqrt{-1}\psi_n^{k}}e^{\sqrt{-1}\varphi_{n+1}^{k}}
\,,\label{identity2:sp}
\end{equation}
with
\begin{eqnarray*}
e^{\sqrt{-1}\psi_n^{k}}=\cos\psi_n^k+\sqrt{-1}\sin \psi_n^k
=\frac{X_{n+1}^k-X_n^k}{a_n}+\sqrt{-1}~\frac{v_{n+1}^k-v_n^k}{a_n}\,,\\
e^{\sqrt{-1}\varphi_n^{k}}=\cos\varphi_n^k+\sqrt{-1}\sin \varphi_n^k=
\frac{X_{n}^{k+1}-X_n^k}{\frac{1}{c_k}}
+\sqrt{-1}~\frac{v_{n}^{k+1}+v_n^k}{\frac{1}{c_k}}\,.
\end{eqnarray*}
We then obtain the following system of two discrete equations
\begin{eqnarray}
&&(X_{n+1}^{k+1}-X_{n+1}^{k}-X_{n}^{k+1}+X_{n}^{k})
(X_{n+1}^{k}-X_{n}^{k+1})\nonumber\\
&&\qquad -
(v_{n+1}^{k+1}+v_{n+1}^{k}-v_{n}^{k+1}-v_{n}^{k})
(v_{n+1}^{k}+v_{n}^{k+1})=0\,,\label{eqn:discreteSP1}\\
&&
(X_{n+1}^{k+1}-X_{n+1}^{k}+X_{n}^{k+1}-X_{n}^{k})
(v_{n+1}^{k+1}-v_{n+1}^{k}-v_{n}^{k+1}+v_{n}^{k})\nonumber\\
&&\qquad -(X_{n+1}^{k+1}+X_{n+1}^{k}-X_{n}^{k+1}-X_{n}^{k})
(v_{n+1}^{k+1}+v_{n+1}^{k}+v_{n}^{k+1}+v_{n}^{k})=0\,.\label{eqn:discreteSP2}
\end{eqnarray}
Note that the second equation (\ref{eqn:discreteSP2}) 
is further simplified to
\begin{equation}
(X_{n+1}^{k}-X_{n}^{k+1})(v_{n+1}^{k+1}+v_{n}^{k})
+(X_{n+1}^{k+1}-X_{n}^{k})(v_{n+1}^{k}+v_{n}^{k+1})=0\,.
\end{equation}
Replacing $X_n^k$ by $X_n^k+\sum_{j=0}^{k-1}\frac{1}{c_j}$, 
(\ref{eqn:discreteSP1}) and (\ref{eqn:discreteSP2}) become 
\begin{eqnarray}
&&(X_{n+1}^{k+1}-X_{n+1}^{k}-X_{n}^{k+1}+X_{n}^{k})
\left(X_{n+1}^{k}-X_{n}^{k+1}-\frac{1}{c_k}\right)\nonumber\\
&&\qquad -
(v_{n+1}^{k+1}+v_{n+1}^{k}-v_{n}^{k+1}-v_{n}^{k})
(v_{n+1}^{k}+v_{n}^{k+1})=0\,,\label{eqn:discreteSP1-2}\\
&&
\left(\frac{2}{c_k}+X_{n+1}^{k+1}-X_{n+1}^{k}+X_{n}^{k+1}-X_{n}^{k}\right)
(v_{n+1}^{k+1}-v_{n+1}^{k}-v_{n}^{k+1}+v_{n}^{k})\nonumber\\
&&\qquad -(X_{n+1}^{k+1}+X_{n+1}^{k}-X_{n}^{k+1}-X_{n}^{k})
(v_{n+1}^{k+1}+v_{n+1}^{k}+v_{n}^{k+1}+v_{n}^{k})=0\,.\label{eqn:discreteSP2-2}
\end{eqnarray}
Note that this form was obtained in \cite{discreteSP} by using the
bilinear method. Taking the continuous limit $c_k\to 0$ of
(\ref{eqn:discreteSP1-2}) and (\ref{eqn:discreteSP2-2}), 
we obtain the semi-discrete short pulse equation 
(\ref{semi-discrete_SP1}) and (\ref{semi-discrete_SP2}).  

\section{$\tau$-function and soliton type solutions}

In this section, we list the $\tau$-function and the bilinear equations which give rise to the
soliton and breather type solutions to the equations and curve motions discussed in Section 2.
Although they have been already discussed in \cite{Inoguchi,Inoguchi:semi,discreteSP}, we collect
and present the results for completeness and the convenience of readers. It should be remarked that
all the solutions can be expressed in terms of one $\tau$-function.


The solutions can be expressed in the following form:
\begin{eqnarray}
&&\theta_{n,l}^{m,k}(s,t,\zeta,y)=\frac{2}{\sqrt{-1}}\log
 \frac{\tau_{n,l}^{m,k}}{\taus_{n,l}^{m,k}}\,,\\
&&\gamma_{n,l}^{m,k}(s,t,\zeta,y)=
\left[\begin{array}{c}
-\frac{1}{2}(\log \tau_{n,l}^{m,k}\taus_{n,l}^{m,k})_y\\[2mm]
\frac{1}{2\sqrt{-1}}
\left(\log
\frac{\tau_{n,l}^{m,k}}{\taus_{n,l}^{m,k}}
\right)_y
\end{array}\right]\,.
\end{eqnarray}
Here, the $\tau$-function $\tau_{n,l}^{m,k}(s,t,\zeta,y)$ is given by~\cite{Inoguchi}:
\begin{equation}
\fl \tau_{n,l}^{m,k}(s,t,\zeta,y) = \exp\left[-\left(s+\zeta+\epsilon l 
+ \sum_{n'}^{n-1}a_{n'} 
+ \sum_{m'}^{m-1}b_{m'}+\sum_{k'}^{k-1}\frac{1}{c_{k'}}\right)y\right]
~\det\left(f_{j-1}^{(i)}\right)_{i,j=1,\ldots,N},\label{Casorati} 
\end{equation}
\begin{equation}
f_{j}^{(i)} = e^{\eta_i} + e^{\mu_i},\label{tau:Casorati_entries0}
\end{equation}
\begin{equation}
\fl
\left\{
\begin{array}{l}
 {\displaystyle  e^{\eta_i}= \alpha_ip_i^{j}
(1-\epsilon p_i)^{-l} 
\prod_{n'}^{n-1}(1-a_{n'}p_i)^{-1}\prod_{m'}^{m-1}(1-b_{m'}p_i)^{-1}
\prod_{k'}^{k-1}\left(1-\frac{c_{k'}}{p_i}\right)^{-1}
e^{p_i s - 4p_i^3 t
+\frac{p_i}{1-\epsilon^2p_i^2}\zeta +  \frac{1}{p_i}y }},\\[4mm]
{\displaystyle e^{\mu_j}=  
\beta_i(-p_i)^{j} (1+\epsilon p_i)^{-l}\prod_{n'}^{n-1}
 (1+a_{n'}p_i)^{-1}\prod_{m'}^{m-1}(1+b_{m'}p_i)^{-1}
\prod_{k'}^{k-1}\left(1+\frac{c_{k'}}{p_i}\right)^{-1}
e^{-p_is  + 4p_i^3 t-\frac{p_i}{1-\epsilon^2p_i^2}\zeta -  \frac{1}{p_i}y }}.
\end{array}\right.
\label{tau:Casorati_entries}
\end{equation}
The parameters are chosen as follows:
\begin{enumerate}
 \item {\bf N-soliton solution:}
\begin{equation}\label{param:soliton}
p_i,\ \alpha_i\in\mathbb{R},\quad \beta_i\in\sqrt{-1}\mathbb{R}\quad (i=1,\ldots,N).
\end{equation}
 \item {\bf M-breather solution:} 
\begin{equation}\label{param:breather}
\begin{array}{l}\medskip
 {\displaystyle  N=2M,\quad p_i,\ \alpha_i,\ \beta_i \in\mathbb{C}\quad (i=1,\ldots,2M),}\\
{\displaystyle p_{2j}=p_{2j-1}^*,\quad \alpha_{2j}=\alpha_{2j-1}^*,\quad  
\beta_{2j}=-\beta_{2j-1}^*\quad (j=1,\ldots,M).}
\end{array}
\end{equation}
\end{enumerate}
The bilinear equations which are necessary to recover the equations and curve motions are given as
follows. Note that we only show the relevant independent variables, and other variables can be
regarded as parameters.

\noindent{\bf Continuous case}\cite{Inoguchi,discreteSP}: $\tau=\tau(s,t,y)$,
\begin{eqnarray}
&& \frac{1}{2}D_sD_y~\tau\cdot\tau = -\left(\tau^*\right)^2,\label{bl1}\\
 && D_s^2~\tau\cdot\taus =0,\label{bl2}\\
 &&\left(D_s^3+D_t\right)~\tau\cdot\taus = 0.\label{bl3}
\end{eqnarray}
\noindent{\bf Semi-discrete case}\cite{Inoguchi:semi,discreteSP}: $\tau=\tau_l(\zeta,y)$,
\begin{eqnarray}
&& D_\zeta~\tau_l\cdot\tau^*_l=\frac{1}{2\epsilon}\left(\tau^*_{l-1}\tau_{l+1}-\tau_{l+1}^*\tau_{l-1}\right),\label{bl4}\\
&& \tau_l\tau^*_l=\frac{1}{2}\left(\tau^*_{l-1}\tau_{l+1}+\tau^*_{l+1}\tau_{l-1}\right),\label{bl5}\\
&& \frac{1}{2}D_\zeta D_y~\tau_l\cdot\tau_l = -\tau^*_{l+1}\tau^*_{l-1},\label{bl6}\\
&& D_y~\tau_{l+1}\cdot\tau_l = -\epsilon\tau^*_{l+1}\tau^*_l.\label{bl7}
\end{eqnarray}
\noindent{\bf Discrete case}\cite{Inoguchi,discreteSP}: $\tau=\tau_n^{m,k}(y)$,
\begin{eqnarray}
 && D_y~\tau_{n+1}^{m,k}\cdot\tau_n^{m,k} = -a_n\tau^*{}_{n+1}^{m,k}\tau^*{}_n^{m,k},\label{bl8}\\
 && D_y~\tau_{n}^{m+1,k}\cdot\tau_n^{m,k} = -b_m\tau^*{}_{n+1}^{m,k}\tau^*{}_n^{m,k},\label{bl9}\\
 && D_y~\tau_n^{m,k+1}\cdot\taus_n^{m,k} = -\frac{1}{c_k}\taus_n^{m,k+1}\tau_n^{m,k},\label{bl10}\\
 &&b_m\tau^*{}_{n}^{m+1,k}\tau_{n+1}^{m,k} - a_n\tau^*{}_{n+1}^{m,k}\tau_{n}^{m+1,k}
+ (a_n-b_m)\tau^*{}_{n+1}^{m+1,k}\tau_n^{m,k}=0.\label{bl11}
\end{eqnarray}
%
\section{Conclusions}
In this paper, we have discretized several 
soliton equations which admit loop type soliton solutions
through the discrete analogues of the hodograph transformations based on the geometry of plane
curves. More concretely, we have constructed semi-discrete and fully discrete versions of the WKI
elastic beam equation, the complex Dym equation and the short pulse equation, and presented the
$\tau$-function which gives rise to the soliton and breather solutions.

Geometric consideration is effective for discretization of soliton equations which admit soliton
solutions with singularities, and it may be also applicable to other soliton equations. For example,
the soliton equations arising from the curve motions in the Minkowski plane are of so-called
``defocusing type'' which have nonlinear terms with different signs compared to the equations
discussed in this paper, and it is known that the structure and behaviour of the solutions are quite
different. The geometric consideration may also be useful to discretize this class of
equations. This problem will be reported in a forthcoming paper.

\appendix
\section*{Appendix: Hodograph transformations}
\setcounter{section}{1}
{\bf The WKI elastic beam equation:}\\
A conservation law of the potential mKdV equation (\ref{eqn:pmKdV}) is
given by 
\begin{equation}
(\cos \theta)_t+\left(\frac{1}{2}(\theta_s)^2\cos \theta 
-\theta_{ss}\sin \theta\right)_s=0\,.\label{pmKdV:conservation1}
\end{equation}
Consider the hodograph transformation~\cite{Ishimori}
\begin{equation}
x(s,t)=\int_0^{s}\cos \theta(s',t) ds'+x_0\,,\quad t'(s,t)=t\,,
\end{equation}
which leads to 
\begin{equation}
\frac{\partial}{\partial s}=
\cos \theta \frac{\partial}{\partial x}\,,\qquad 
\frac{\partial}{\partial t}=
\frac{\partial}{\partial t'}+
\left(\theta_{ss}\sin \theta-\frac{1}{2}(\theta_s)^2\cos \theta \right)
\frac{\partial}{\partial x}\,.\label{appendix:hodograph-wki}
\end{equation}
Applying (\ref{appendix:hodograph-wki}) to (\ref{pmKdV:conservation1}), 
we obtain 
\begin{equation}
\theta_{t'}+\cos^2\theta 
(\sin \theta)_{xxx}=0\,, 
\end{equation}
which can be rewritten as 
\begin{equation}
(\tan \theta)_{t'}+(\sin \theta)_{xxx}=0\,.\label{wki:ishimori} 
\end{equation}
Introducing a new dependent variable
$v(s,t)=\int_0^s\sin \theta(s',t) ds'+v_0$ (note $\tan \theta=v_x$ and 
$\sin \theta=v_x/\sqrt{1+(v_x)^2}$), 
(\ref{wki:ishimori}) is transformed to
\begin{equation}
v_{t'x}+\left(\frac{v_x}{\sqrt{1+(v_x)^2}}\right)_{xxx}=0\,,
\end{equation}
which is the (potential) WKI elastic beam equation (\ref{eqn:pWKI}).\\
\quad \\
{\bf The complex Dym equation:}\\
A conservation law of the potential mKdV equation (\ref{eqn:pmKdV}) is
given by 
\begin{equation}
(e^{\sqrt{-1}\theta})_t+\left(\frac{1}{2}(\theta_s)^2e^{\sqrt{-1}\theta} 
+\sqrt{-1}\theta_{ss}e^{\sqrt{-1}\theta}
\right)_s=0\,.\label{pmKdV:conservation2}
\end{equation}
Consider the hodograph transformation~\cite{Kawamoto,Dmitrieva}
\begin{equation}
z(s,t)=\int_0^s e^{\sqrt{-1}\theta(s',t)} ds'+z_0\,,\quad t'(s,t)=t\,,
\end{equation}
which leads to 
\begin{equation}
\frac{\partial}{\partial s}=
e^{\sqrt{-1}\theta} \frac{\partial}{\partial z}\,,\qquad 
\frac{\partial}{\partial t}=
\frac{\partial}{\partial t'}+
\left(-\frac{1}{2}(\theta_s)^2e^{\sqrt{-1}\theta} 
-\sqrt{-1}\theta_{ss}e^{\sqrt{-1}\theta}
\right)
\frac{\partial}{\partial z}\,.\label{appendix:hodograph-dym}
\end{equation}
Applying (\ref{appendix:hodograph-dym}) to (\ref{pmKdV:conservation2})
and introducing a new dependent variable $r=e^{\sqrt{-1}\theta}$, 
we obtain the complex Dym equation
\begin{equation}
r_{t'}+r^3 
(r)_{zzz}=0\,.
\end{equation}
Note that 
\begin{equation}
z=x+\sqrt{-1}v\,,\qquad r=\frac{\partial z}{\partial s}\,.
\end{equation}
\quad \\
{\bf The short pulse equation:}\\
A conservation law of the sine-Gordon equation (\ref{eqn:sG}) is
given by 
\begin{equation}
(\cos \theta)_y+\left(\frac{(\theta_y)^2}{8}\right)_s=0
\,.\label{sG:conservation}
\end{equation}
Consider the hodograph transformation~
\cite{Matsuno-SP,Sakovich,Sakovich2,Rogers}
\begin{equation}
x(s,y)=\int_0^s \cos \theta(s',y) ds'+x_0\,,\quad y'(s,y)=y\,,
\end{equation}
which leads to 
\begin{equation}
\frac{\partial}{\partial s}=
\cos \theta \frac{\partial}{\partial x}\,,\qquad 
\frac{\partial}{\partial y}=
\frac{\partial}{\partial y'}-
\frac{(\theta_y)^2}{8}
\frac{\partial}{\partial x}\,.\label{appendix:hodograph-SP}
\end{equation}
Introduce a new dependent variable 
\begin{equation}
v(s,y)=\int_0^s\sin \theta (s',y) ds'+v_0=\int_0^s\frac{\theta_{ys'}(s',y)}{4}ds'+v_0=\frac{1}{4}\theta_y\,,
\end{equation}
then it follows 
\[
v_x=\tan \theta\,. 
\]
Applying (\ref{appendix:hodograph-SP}) to (\ref{sG:conservation}), 
we obtain 
\begin{equation}
\left(\frac{\partial}{\partial y'}-2v^2\frac{\partial}{\partial x}\right)
\cos \theta=-4vv_x\cos \theta\,.
\end{equation}
this can be rewritten as 
\begin{equation}
\left(\frac{\partial}{\partial y'}-2v^2\frac{\partial}{\partial x}\right)
\frac{1}{\cos^2 \theta}=8vv_x\frac{1}{\cos^2 \theta}\,.
\end{equation}
From 
\[
\frac{1}{\cos^2\theta}=1+\tan^2\theta=1+v_x^2\,, 
\]
it follows that 
\begin{equation}
\left(\frac{\partial}{\partial y'}-2v^2\frac{\partial}{\partial x}\right)
(1+v_x^2)=8vv_x(1+v_x^2)\,,
\end{equation}
which is nothing but the short pulse equation
\begin{equation}
 v_{xy'} = 4v + \frac{2}{3}\left(v^3\right)_{xx}\,.
\end{equation}


\end{document}